\documentclass[aps,twocolumn,prd,
showpacs,showkeys,
preprintnumbers,superscriptaddress,
nobibnotes,floatfix,longbibliography,notitlepage,nofootinbib
]{revtex4-1}

\pdfoutput=1
\usepackage{amsmath}
\usepackage{amsfonts}
\usepackage{amssymb}
\usepackage{mathrsfs}
\usepackage{graphicx}
\usepackage{color}
\usepackage{slashed}

\usepackage{hyperref}
\usepackage{multirow}
\usepackage{upgreek}
\usepackage[capitalise]{cleveref}

\newcommand{\ie}{{\it i.e.}}

\newcommand{\eg}{{\it e.g.}}

\begin{document}
\preprint{IPMU22-0010}

\title{Hadrophilic Light Dark Matter from the Atmosphere}

\author{Carlos A. Arg{\"u}elles}
\email{carguelles@fas.harvard.edu}
\affiliation{Department of Physics \& Laboratory for Particle Physics and Cosmology, Harvard University, Cambridge, MA 02138, USA}

\author{V\'ictor Mu\~noz}
\email{vicmual@ific.uv.es}
\affiliation{Instituto de F\'isica Corpuscular, Universidad de Valencia and CSIC, Edificio Institutos Investigac\'ion, Catedr\'atico Jos\'e Beltr\'an 2, 46980 Spain}

\author{Ian M. Shoemaker} 
\email{shoemaker@vt.edu}
\affiliation{Center for Neutrino Physics, Department of Physics, Virginia Tech, Blacksburg, VA 24061, USA}

\author{Volodymyr Takhistov} \email{volodymyr.takhistov@ipmu.jp}
\affiliation{Kavli Institute for the Physics and Mathematics of the Universe (WPI), UTIAS \\The University of Tokyo, Kashiwa, Chiba 277-8583, Japan}
\date{\today}

\begin{abstract}
Light sub-GeV dark matter (DM) constitutes an underexplored target, beyond the optimized sensitivity of typical direct DM detection experiments.
We comprehensively investigate hadrophilic light DM produced from cosmic-ray collisions with the atmosphere.
The resulting relativistic DM, originating from meson decays, can be efficiently observed in variety of experiments, such as XENON1T.
We include for the first time decays of $\eta$, $\eta^{\prime}$ and $K^+$ mesons, leading to
improved limits for DM masses above few hundred MeV.
We incorporate an exact treatment of the DM attenuation in Earth and demonstrate that nuclear form factor effects can significantly impact the resulting testable DM parameter space. Further, we establish projections for upcoming experiments, such as DARWIN, over a wide range of DM masses below the GeV scale.
\end{abstract}
\maketitle

\section{Introduction}

Extensive astrophysical observations point towards the existence of dark matter (DM), currently constituting $\sim85\%$ of matter in the Universe~(see~\cite{Bertone_2005,Gelmini:2015zpa} for a review).
Despite decades of searching for its possible non-gravitational interactions, the nature of the DM remains elusive.
One extensively explored paradigm is that of Weakly Interacting Massive Particles (WIMPs), where the DM has a typical mass in the range of 1~GeV-100~TeV, and often appears in theories aiming to address the hierarchy problem.
A less studied scenario, which could also appear in variety of well-motivated models~(see \textit{e.g.}~\cite{Feng:2008ya,Boehm:2003hm,Lin:2011gj,Lin:2011gj,Hochberg:2014dra,Hochberg:2014kqa}), is sub-GeV (light) DM.
This later case is the focus of this article.

\begin{figure*}[t]
    \centering
    \includegraphics[width=\columnwidth]{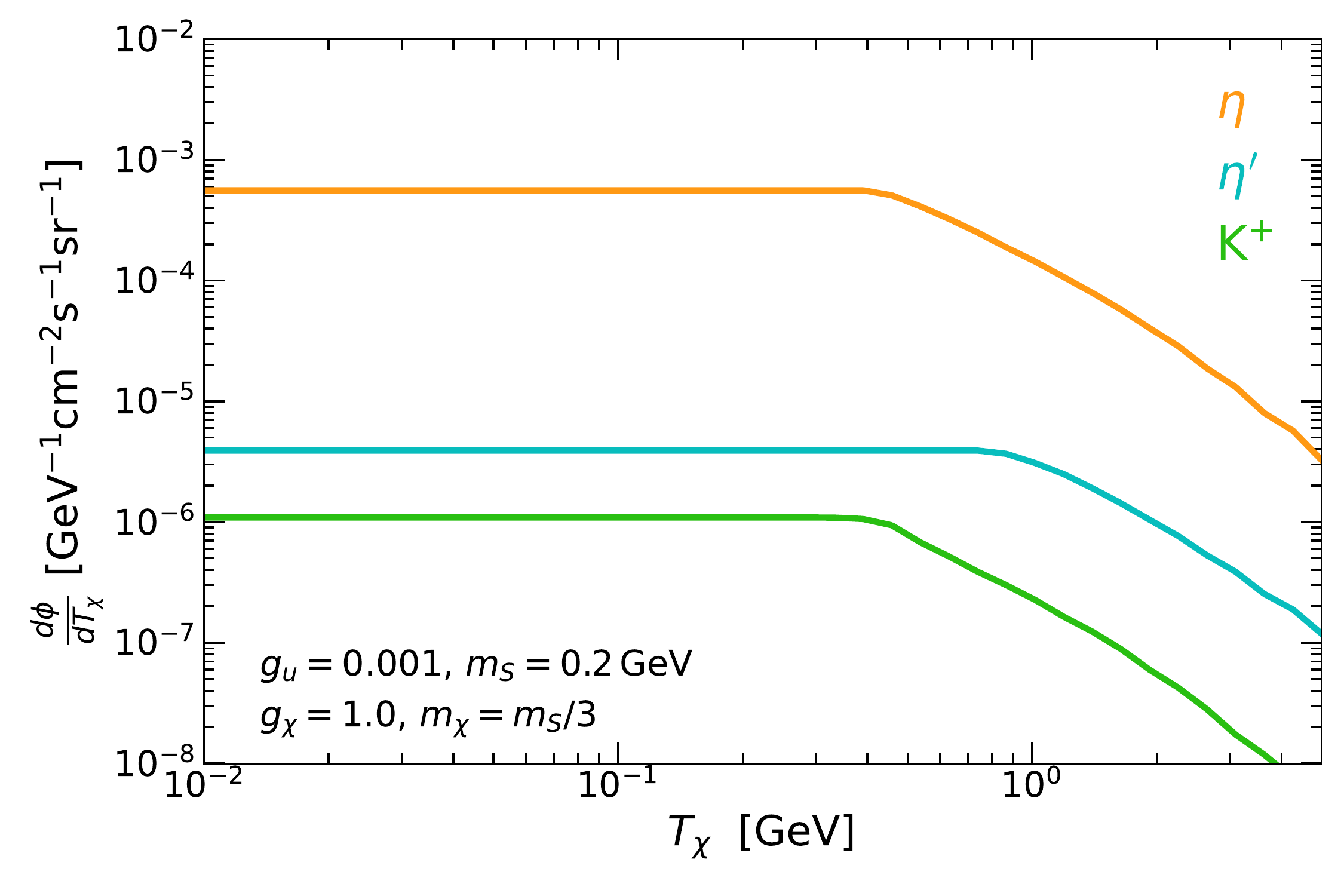}
    \hspace{3mm}
    \includegraphics[width=\columnwidth]{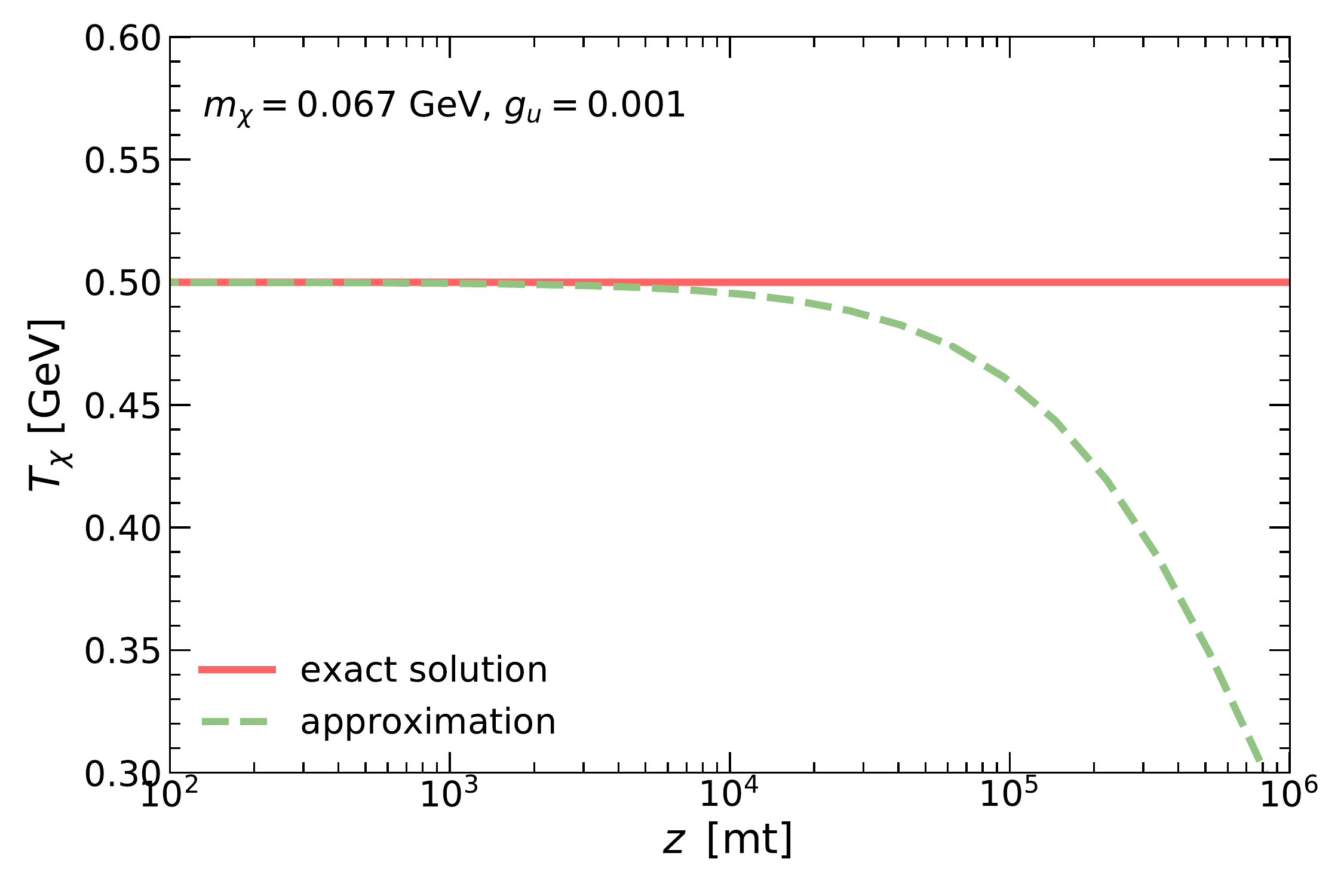}
    \caption{\textbf{Left:} DM flux spectrum at surface of the Earth in terms of DM energy $T_{\chi}$, produced from atmospheric decays of mesons. \textbf{Right:} DM energy loss due to attenuation as it propagates through the Earth, as a function of the depth. The exact solution of Eq.~\eqref{eq:attenuation} (solid red curve) and the expected loss in approximate treatment of Ref.~\cite{Alvey:2019zaa} (dashed green curve), which does not incorporate the nuclear form factor, are displayed.}
    \label{fig:dmflux}
\end{figure*}

A cornerstone of DM searches are direct DM detection experiments, which looks for energy depositions on a target material associated with interactions within the experiment of traversing DM from the Milky Way halo.
Conventional ton-scale direct DM detection experiments with keV-level energy thresholds~(see \textit{e.g.}~\cite{XENON:2017vdw,LUX:2016ggv}) face significant challenges to search for light DM, since it is not expected to produce sufficient nuclear recoil when DM interacts with the target material.
 
Observational signatures of DM could be drastically distinct from nominal expectations if the DM speed distribution contains a significant relativistic component.
Such component can be naturally realized when DM is produced in cosmic-rays collisions with the atmosphere.
The ``atmospheric beamdump'' has been historically exploited as an essential instrument for exploring the properties of neutrinos, contributing to the discovery of neutrino oscillations~\cite{Super-Kamiokande:1998kpq}.
Recent studies have highlighted that the atmospheric beamdump is a unique tool to explore new physics beyond the Standard Model (SM), including long-lived particles~\cite{Arguelles:2019ziu}, heavy neutral leptons~\cite{Coloma:2019htx}, millicharge particles~\cite{Plestid:2020kdm,Harnik:2020ugb,ArguellesDelgado:2021lek}, particle DM~\cite{Alvey:2019zaa,Su:2020zny}, supersymmetric particles~\cite{Candia:2021bsl}, and magnetic monopoles~\cite{Iguro:2021xsu}. 
This natural beam dump produces a steady flux that can be exploited by terrestrial experiments.
Other possible sources of a relativistic DM component include boosted DM~(see \textit{e.g.}~\cite{Agashe:2014yua}), DM accelerated in astrophysical sources~\cite{Hu:2016xas,Dunsky:2018mqs,Li:2020wyl}, the result of cosmic-ray DM scattering~\cite{Dent:2019krz,Cappiello:2018hsu,Bringmann:2018cvk,Ema:2018bih,Dent:2020syp,Ema:2020ulo} or evaporation of primordial black holes~\cite{Calabrese:2021src}.
These DM components are distinct from conventional halo DM analyses~(\textit{e.g.}~\cite{DelNobile:2013cva,Chen:2021qao}).

In this article, we comprehensively analyze light hadrophilic DM produced in the atmosphere by cosmic-ray collisions.
In particular, we focus on hadrophilic dark sectors, which have been discussed within variety of scenarios (see \textit{e.g.}~\cite{Batell:2017kty,Batell:2018fqo,Batell:2021snh,Elor:2021swj}).
We improve upon earlier work~\cite{Alvey:2019zaa} of hadrophilic DM producing in an atmospheric beamdump in many significant aspects which include: considering a more realistic model of the atmospheric flux, including of additional mesons whose deay produces DM, and perform a proper treatment of the attenuation of DM through the Earth.
We also present projections of this scenario for upcoming and proposed experiments.

\section{Hadrophilic Light Dark Matter}

Collisions of the almost isotropic cosmic-ray flux, primarily composed of protons, with the atmosphere results in copious production of Standard Model particles from GeV to well over TeV energies~\cite{Zyla:2020zbs}.
This is directly analogous to a fixed target beam dump experiment.
The atmospheric beam dump can produce particles associated with physics beyond the SM. 

To explore sub-GeV DM, we consider a minimal realization within a framework of flavor-specific, hadrophilic, scalar mediator~\cite{Batell:2017kty,Batell:2018fqo}.
The effective Lagrangian includes
\begin{align} \label{eq:model}
    \mathcal{L} \supset &~ i \overline{\chi}(\slashed{D} - m_{\chi})\chi + \dfrac{1}{2}\partial_{\mu}S\partial^{\mu}S - \dfrac{1}{2}m_S^2 S^2 \notag\\
    & - \Big(g_{\chi} S \overline{\chi}_L \chi_R + g_u S \overline{u}_L u_R + h.c.\Big),
\end{align}
where $S$ is a singlet scalar mediator with mass $m_S$, $\chi$ is a singlet Dirac fermion DM with mass $m_{\chi}$ stabilized by a $Z_2$ symmetry, and $g_u$ and $g_{\chi}$ are effective couplings.
Here, the scalar mediates the interactions between the dark sector and the visible sector, in our work, the up quarks.

In the medium where the DM propagated and in the experiments instrumented volume, DM will scatter against a nucleus with differential cross-section given by 
\begin{align}
\frac{d\sigma_{\chi N}}{dT_r} =&~ \frac{g_{\chi}^{2}(Z y_{spp}+(A-Z)y_{snn})^2}{8\pi}\frac{(T_r+2m_{N})}{(T_{\chi}^{2}+2m_{\chi}T_{\chi})} \notag\\
&~\times \dfrac{(2m_{\chi}^2+m_{N}T_r)}{
(2m_{N}T_r+m_{S}^2)^2}F_{H}^{2}(2m_{N}T_r),  
\end{align}
where $A$, $T_r$, and $m_{N}$ are the nucleon number, recoil energy, and nucleus mass, respectively.
The term $F_{H}$ is the nuclear Helm form factor~\cite{Helm:1956} (see also \textit{e.g.}~\cite{Duda:2006uk})
\begin{equation} \label{eq:helmform}
    F_H(q^2) = \dfrac{3 j_1(q R_1)}{q R_1} e^{-\frac{1}{2}q^2 s^2}~,
\end{equation}
where $q = \sqrt{2 m_N T_r}$ is the momentum transfer, $j_1(x)$ is a spherical Bessel function of the first kind, $R_1 = \sqrt{R_A^2 - 5 s^2}$ with $R_A \simeq 1.2 A^{1/3}$~fm and $s \simeq 1$~fm.
The $g_{\chi}$ is the scalar-DM coupling, which we fix at 1.0, while the effective scalar-nucleon couplings, $y_{spp}$, $y_{npp}$ are given by~\cite{Batell:2018fqo}
\begin{eqnarray}
&y_{spp}=0.014\, g_{u}\dfrac{m_p}{m_u},  \\ \nonumber
&y_{snn}=0.012\, g_{u}\dfrac{m_n}{m_u},
\end{eqnarray}
with $m_p$, $m_n$, and $m_u$ the masses of the proton, neutron, and up-quark.

The nuclear interaction cross-section can be related to the conventional spin-independent cross-section, $\sigma^{\rm SI}$, by
\begin{equation} \label{eq:sixsec}
    \sigma_{\chi N} =  \sigma^{\rm SI} A^2 \Big(\dfrac{m_N}{m_p} \dfrac{(m_{\chi} + m_p)}{(m_{\chi} + m_N)}\Big)^2.
\end{equation}

\begin{figure*}[t]
    \centering
    \includegraphics[width=\columnwidth]{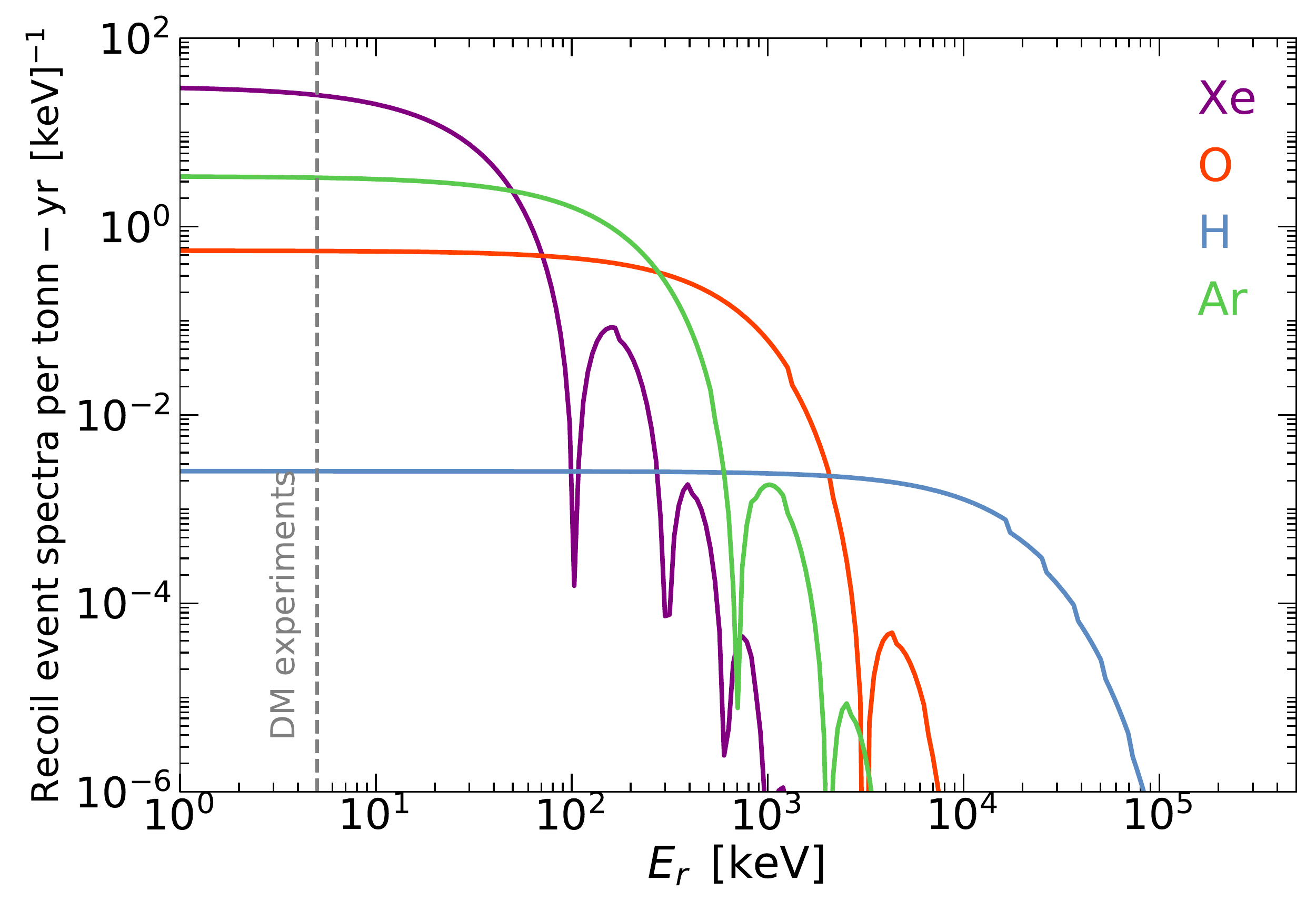}
    \hspace{3mm}
    \includegraphics[width=\columnwidth]{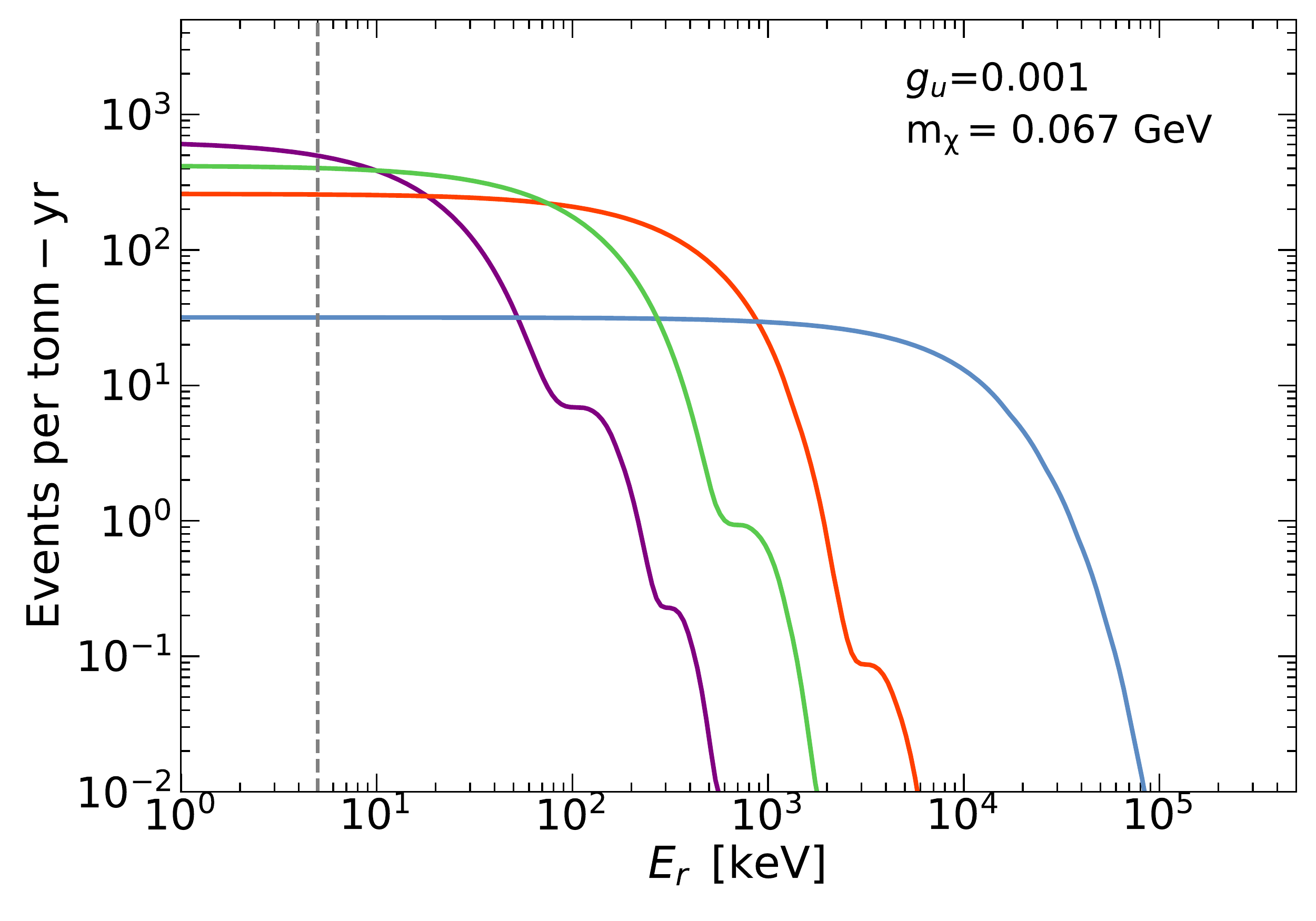}
\caption{
\textbf{Left:} Differential event spectra for different target materials as a function of nuclear recoil energy $E_r$, including xenon (Xe, purple), oxygen (O, orange) and hydrogen (H, blue). The total exposure is one ton-year. \textbf{Right:} Total integrated number of recoil events above given experimental energy threshold $E_r$, for xenon (purple), oxygen (orange) and hydrogen (blue). The considered model values are $g_u = 10^{-3}$ and $m_{\chi} = 0.067$~GeV.}
\label{fig:Recoils}
\end{figure*}

\section{Atmospheric dark matter production}

The scalar mediator $S$ can be abundantly produced in meson decays in cosmic-ray air showers.
The production is modeled by the cascade equation~(\eg~\cite{Gondolo:1995fq}), which, omitting the interaction terms, contains the following source term for DM production
\begin{align} \label{eq:sproduction}
{ d\Phi_S\over dE_{S}\,d\Omega\,dX} =~& \sum_{M} \int d E_M \Big(\dfrac{1}{ \rho(X)\lambda_M(E_M)} \notag\\
&~\times \dfrac{d \Phi_M}{dE_{M} \,d\Omega}(E_M,\cos\theta,X)\,    \\
&~\times\dfrac{d n }{d E_{S}}( E_{M}, E_S)\Big), \notag 
\end{align} 
where the sum $M$ runs over all possible mesons that can decay to $S$, $\rho$ is the density of the atmosphere, $X$ is the column depth (\ie~the depth in g/cm$^2$ measured from the top of the atmosphere to production point), and $\lambda_{M} \equiv \gamma_{M}\beta_{M}c\tau_{M} $ is the laboratory decay length of the meson that includes the boost factor $\gamma_{M}\beta_{M}$ and its proper lifetime $\tau_{M}$.
Here, the differential production rate of mesons in the shower per unit of solid angle is given by $(d \Phi_{M}/dE_{M} \,d\Omega)$. 

In Eq.~\eqref{eq:sproduction}, the number of scalar mediators with energies between $E_{S}$ and $E_{S}+dE_{S}$ produced in the decay of the meson $M$ is given by $(d n/d E_S)$.
The explicit expression for the last quantity depends on whether the scalar is produced in two-body or three-body decays.
We focus on the simplest case involving only two-body decays of mesons.
In this case, the energy distribution in the laboratory frame is given by
\begin{align}
\frac{dn}{d E_S}= \frac{ \mathrm{Br}(M \rightarrow S + P )} 
{ p_M \sqrt{ \mathcal{K} \left(1,\frac{m_{S}^2}{m_{M}^{2}}, \frac{m_{p}^2}{m_{M}^2}\right) }
},
\label{eq:distribution}
\end{align}
where $\mathrm{Br}(M \rightarrow S + P)$ is the meson branching fraction to the scalar and an additional particle $P$, $\mathcal{K}$ is the Källen function~\cite{Kallen:1964lxa}, and $p_{M}$ is the meson momentum.

For scalar mediators sufficiently strongly coupled to the DM, they will promptly decay to a pair of $\chi\bar{\chi}$ in the particle shower.
The resulting DM production rate can be obtained using mediator production of Eq.~\eqref{eq:sproduction} via
\begin{equation}
 {d\Phi_{\chi}\over dE_{\chi}\,d\Omega\,dX}  = \int d E_{S}  { d\Phi_S\over dE_{S}\,d\Omega\,dX} \, {d n \over d E_{\chi}}( E_{S}, E_{\chi}).
\label{eq:DMproduction}
\end{equation}

Eq.~\eqref{eq:DMproduction} implies that the DM is produced at the same point (height) as the scalar mediator, while accounting for the energy distribution of the DM in the decay.
If the coupling between the scalar and the dark matter is not strong enough, the mediator could be made a long-lived particle and the production of DM would be suppressed.
Integrating Eq.~\eqref{eq:DMproduction} over the column depth yields the expected DM flux at the surface of the Earth per unit of solid angle and per energy bin.

In this work, we include for the first time the production of scalar mediator from the decays of $\eta$, $\eta^{\prime}$ as well as charged kaons in the atmosphere, via $\eta \rightarrow \pi^0 S$, $\eta^{\prime}\rightarrow \pi^0 S$ as well as $K^+ \rightarrow \pi^+ S$ channels\footnote{This is in contrast with previous work~\cite{Alvey:2019zaa}, which only considered $\eta$ decays. Different meson decay channels can themselves be used to investigate new physics~(e.g.~\cite{Escribano:2016ntp,Escribano:2020jvu}).}.
We review the branching ratios for these processes in Appendix~\ref{sec:scalarmed}. 
We numerically solve for the production rate of the mesons in the atmospheric cascade using the \texttt{Matrix Cascade Equation} (\texttt{MCEq}) software package~\cite{Fedynitch:2015zma,Fedynitch:2012fs}, which includes several models for the cosmic-ray spectrum, hadronic interactions, and atmospheric density profiles.
For our base results we employed the \texttt{SYBILL-2.3} hadronic interaction model~\cite{Fedynitch:2018cbl}, the Hillas-Gaisser cosmic-ray model $H3a$~\cite{Gaisser:2011cc}, and the \texttt{NRLMSISE-00} atmospheric model~\cite{Picone:2002go} to obtain the production rate of mesons, which have been extracted following the procedure of Ref.~\cite{Arguelles:2019ziu,Coloma:2019htx}. 

The resulting DM flux at the surface of the Earth is shown in Fig.~\ref{fig:dmflux}.
We show each of the progenitor mesons $\eta$, $\eta^{\prime}$ and $K^+$, considering $g_u = 10^{-3}$, $g_{\chi} = 1$, $m_S = 0.2$~GeV, and $m_{\chi} = m_S/3$.
With a mass of $m_{\eta^{\prime}} = 957.8$~MeV, $\eta^{\prime}$ allows for additional kinematic regimes compared to $\eta$ and $K^+$, since these have smaller masses; namely $m_{\eta} = 547.9$~MeV and $m_{K^+} = 493.7$~MeV, respectively.

\begin{figure*}[t]
    \centering
    \includegraphics[width=\columnwidth]{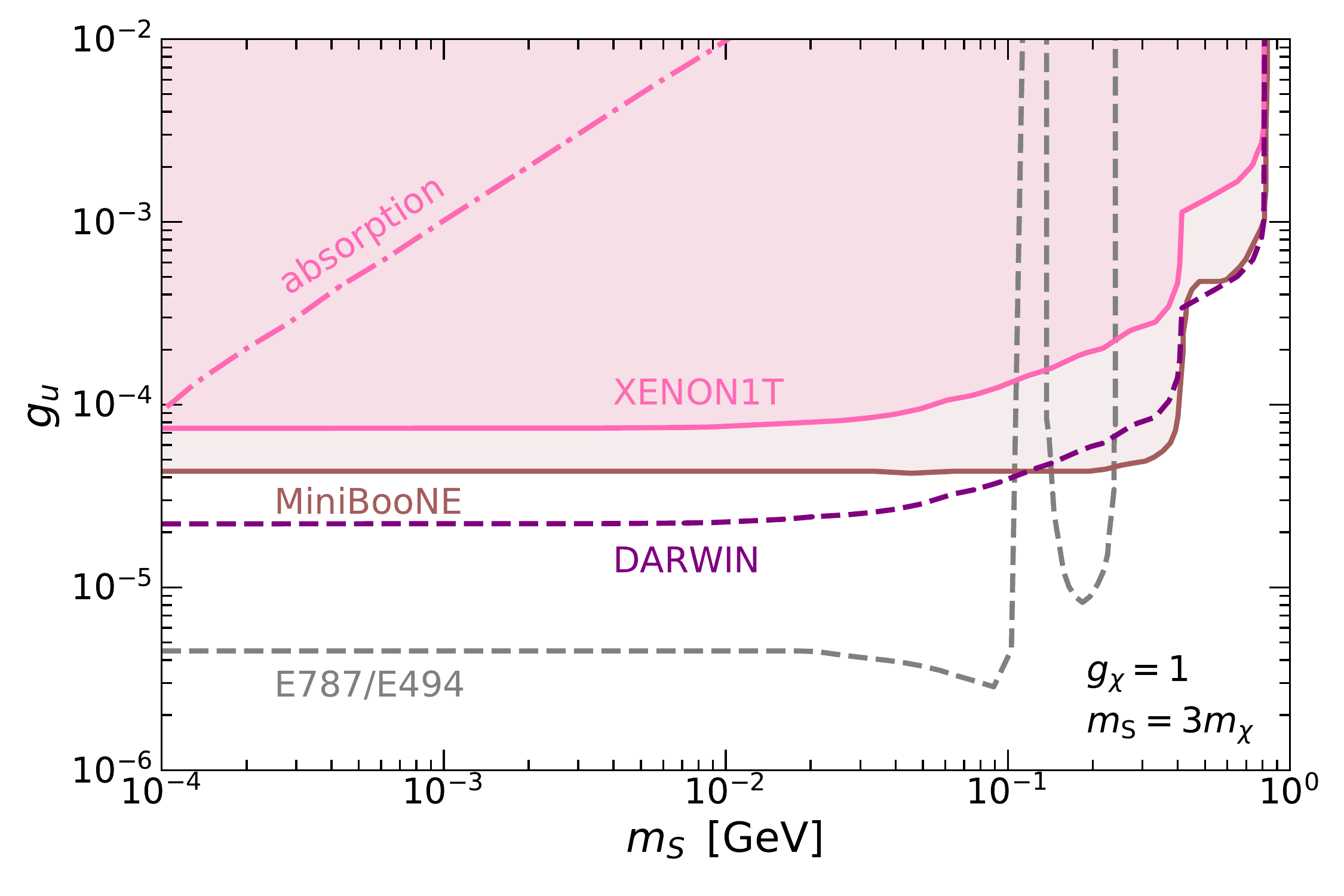}
    \hspace{3mm}
    \includegraphics[width=\columnwidth]{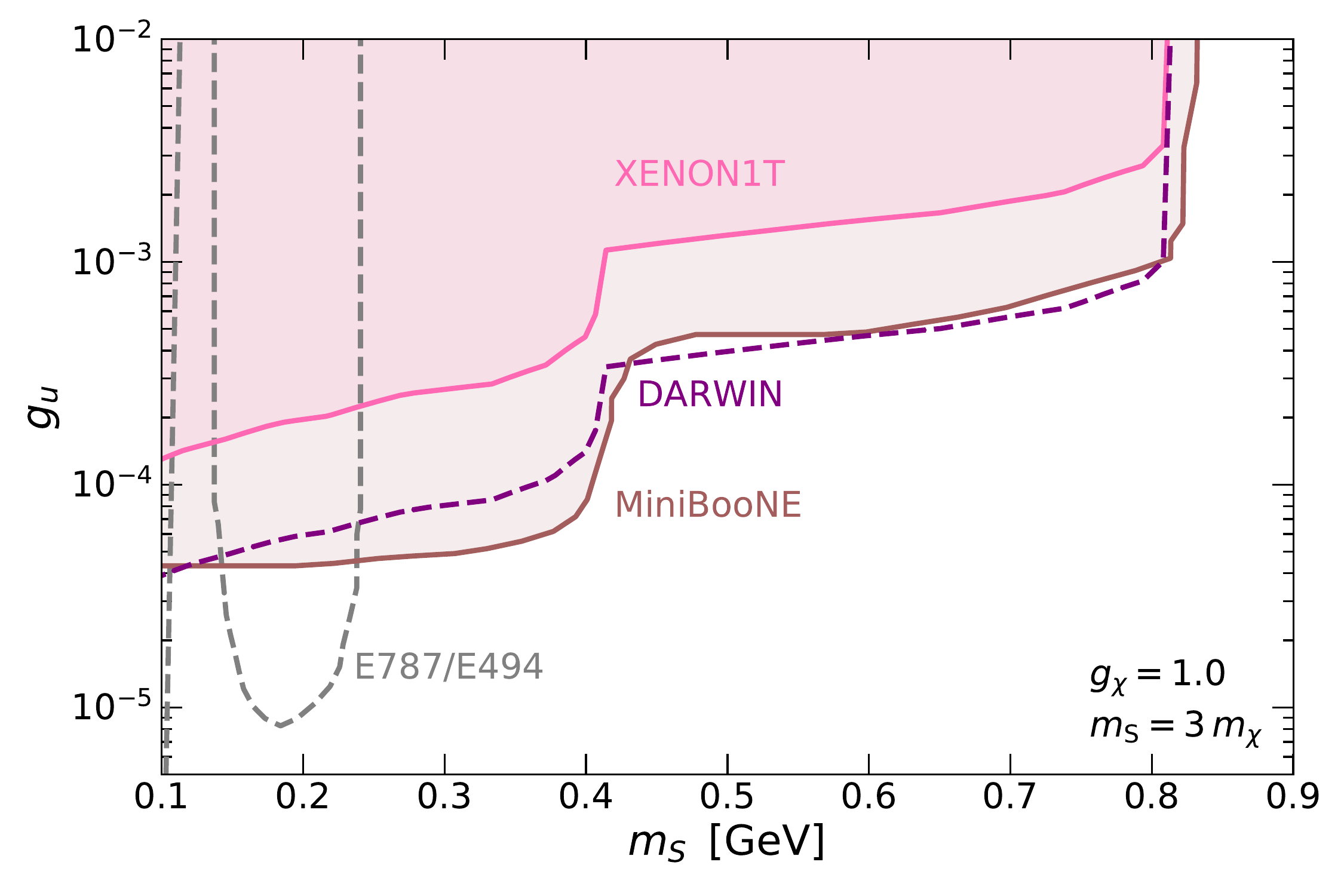}
    \caption{\textbf{Left:} Limits from atmospheric DM flux from meson decays (red), for a hadrophilic scalar mediator of mass $m_S$ on up-quark coupling $g_u$, considering $g_{\chi} = 1$ and $m_S = 3 m_{\chi}$. \textbf{Right:} Same, with focus on parameter space for $m_s$ above 0.1 GeV.}
    \label{fig:limits}
\end{figure*}

\section{Dark Matter Flux Attenuation}

In analogy to SM particles such as muons, the flux of DM with sizable interaction strength is attenuated as it traverses the medium from the top of the atmosphere towards the experiment.
At large cross-sections the medium becomes ``optically thick,'' causing direct DM detection experiments to dramatically lose their sensitivity in this part of the DM parameter space~(e.g.~\cite{Mack:2007xj,Emken:2018run}).

The DM energy loss is described by
\begin{align} \label{eq:attenuation}
\frac{dT_{\chi}}{dz}=-\sum_{N} n_{N}\int_{0}^{T_{r}^{\rm max}}\, T_{r}\,\frac{d\sigma_{\chi N}}{dT_{r}}\,dT_{r},   
\end{align} 
where $T_r$ is the energy lost by a DM particle in a collision with nucleus $N$, and
\begin{align}
T_{r}^{\rm max}=\frac{T_{\chi}^{2}+2m_{\chi}T_{\chi}}{T_{\chi}+(m_{\chi}+m_{N})^{2}/2m_{N}}~.    
\end{align}
To account for DM flux attenuation, we numerically solve Eq.~\eqref{eq:attenuation}.

Our analysis takes into account the relevant effects of the nuclear form factor of Eq.~\eqref{eq:helmform}, which in contrast have been neglected in previous work employing analytic approximation of Eq.~\eqref{eq:attenuation} to treat attenuation~\cite{Alvey:2019zaa}.
The effects of the nuclear form factor in attenuation can significantly impact which DM parameter space regimes are direct DM detection experiments are sensitive to, as has been recently highlighted in~Ref.~\cite{Xia:2021vbz}.

In Fig.~\ref{fig:dmflux} we display the DM flux attenuation using the exact numerical treatment of Eq.~\eqref{eq:attenuation} as well as analytic approximation given in Ref.~\cite{Alvey:2019zaa}.
We highlight that the approximate treatment did not include the effects arising from the nuclear form factor, which suppress large momentum transfer regimes resulting in an overestimation of the DM energy losses for underground experiments lying at a depth below $\sim 1$~km of rock overburden.
Having found $T_{\chi}(z)$ with the procedure outlined above, we can determine the resulting atmospheric DM flux as a function of depth $d \Phi_{\chi} / dT(z)$.

\section{Experimental Constraints}

The differential event rate for a detector with $N_t$ nucleons is given by
\begin{eqnarray}
\frac{dN}{dT_r}=N_{t}\int_{E_{\chi}^{\rm min}} dE_{\chi} \,\epsilon(T_r)\frac{d\sigma_{\chi N}}{dT_r}{ d\Phi_{\chi}\over dT_{\chi}},
\label{eq:event_rate}
\end{eqnarray}
where $\epsilon(T_r)$ is the detection efficiency taken from reference \cite{XENON:2018voc}.
In Fig.~\ref{fig:Recoils} we show the event spectra as a function of the nuclear recoil energy $T_r$ for different target materials, including xenon (Xe), oxygen (O) and hydrogen (H).
The bumps in the spectrum follow the shape of the nuclear Helm factor in the interaction cross-section.
We also display in Fig.~\ref{fig:Recoils} the total integrated number of nuclear recoil events above experimental threshold $T_r$, for different materials and considering characteristic model parameter values of $g_u = 10^{-3}$ and $m_{\chi} = 0.067$~GeV.
These results highlight the importance of low $\mathcal{O}(1)$keV experimental thresholds available for large DM experiments, in accord with the literature results for other physics analyses and complementarity studies~(\eg~\cite{Gelmini:2018ogy,Gelmini:2018gqa,Raj:2019wpy,Munoz:2021sad}).
We note that hadrophilic DM could also induce interactions in large neutrino experiments.
However, such interactions require significantly higher thresholds and the resulting event rates are expected to be suppressed even taking into account the larger available fiducial mass.

\begin{figure}[t]
    \centering
    \includegraphics[width=\columnwidth]{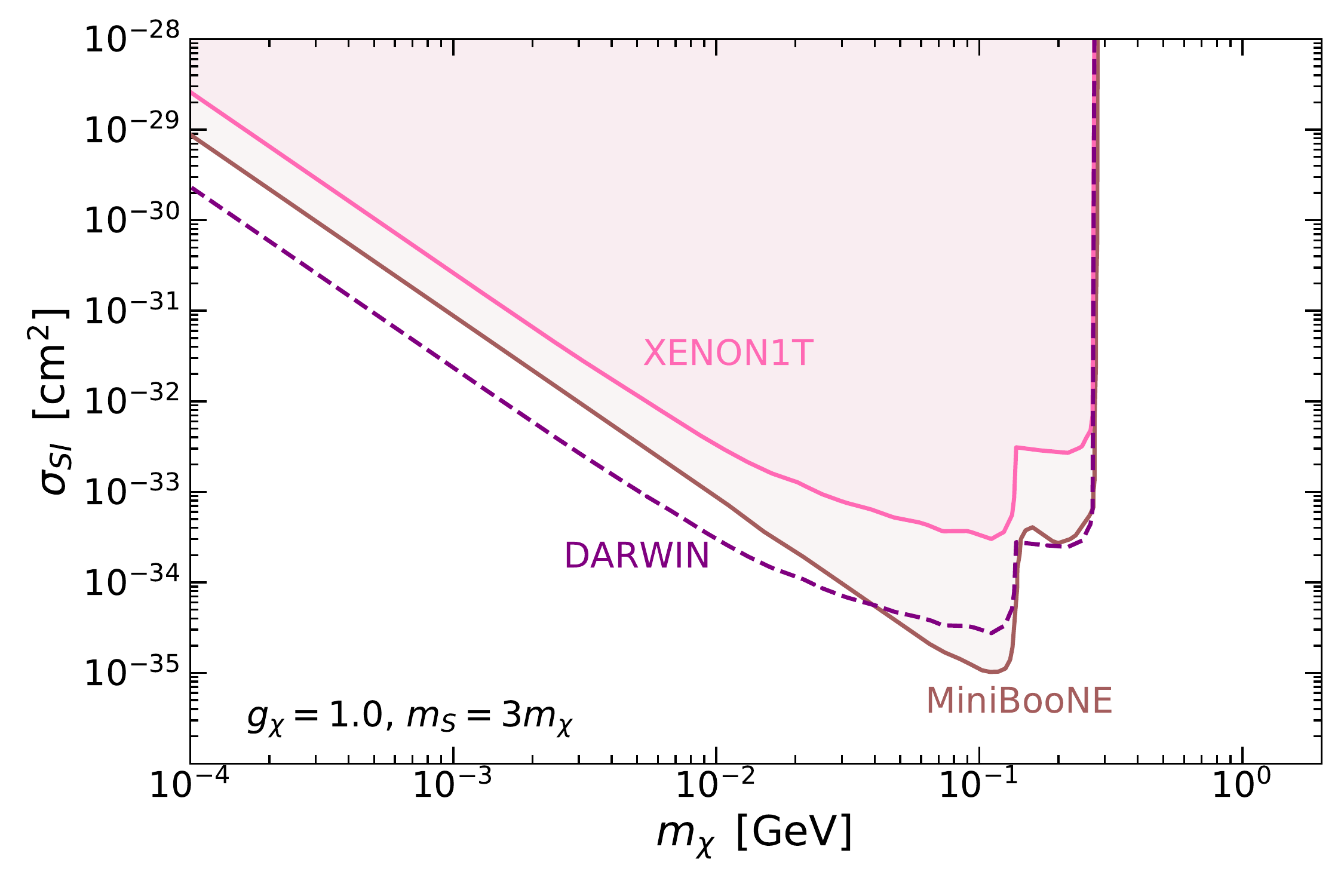}
    \caption{Limits on the spin-independent DM nucleon cross-section $\sigma^{\rm SI}$ as a function of the DM mass $m_{\chi}$.}
    \label{fig:limitsxsec}
\end{figure}
 
To set limits on the model up-quark coupling $g_u$ from XENON1T, we integrate Eq.~\eqref{eq:event_rate} in the recoil energy window with a threshold of 5~keV and a maximum recoil value of 40.9~keV.
Following Ref.~\cite{XENON:2018voc}, we set a limit by requiring to observe more than 3.56 events for an exposure of 278.8 live-days.
The limit in the plane $g_{u}-m_{S}$ is shown in Fig.~\ref{fig:limits}. 
For comparison, we also display constraints on the model from E787/E949 and MiniBooNE experiments following Ref.~\cite{Batell:2018fqo}.
We also display the sensitivity reach projections for the future upgrade of XENON1T, the xenon-based DARWIN experiment~\cite{DARWIN:2016hyl}, which is expected to have a size that is about 50 times larger with a fiducial mass of $\sim 50$ tons.

The effects of implementing an exact numerical treatment of attenuation that also includes the nuclear form factor are shown in Fig.~\ref{fig:limits} with the line ``absorption,'' above which DM flux would be attenuated if the approximate treatment has been used instead.
This further highlights the importance of attenuation for identifying and exploring the optimal DM parameter space with direct DM experiments.    
Inclusion of additional mesons allows us to set new limits above few hundred MeV, as shown on Fig.~\ref{fig:limits}.
This is in contrast to earlier studies of Ref.~\cite{Alvey:2019zaa} that focused solely on $\eta$ and hence were not sensitive to this regime. Depending on the details of the full model as well as cosmological history, additional constraints could become relevant in the parameter space of interest, such as those from Big Bang nucleosynthesis and cosmic microwave background (see e.g. \cite{Batell:2018fqo,Sabti:2019mhn}). We stress, however, that our analysis results are independent of cosmology and the mechanism of DM production.
 
The limits found for the parameters on the model and up-quark coupling $g_u$ can be translated into a stringent constraint for the spin-independent DM cross-section using Eq.~\eqref{eq:sixsec}.
The results are shown in Fig.~\ref{fig:limitsxsec} for DM masses below the GeV scale.
This highlights the significance of atmospheric DM production for exploration of light sub-GeV DM.

\section{Conclusions}

Light sub-GeV DM is a promising target for exploration beyond the traditional WIMP paradigm. However, the associated DM parameter space is challenging to probe with conventional direct DM detection experiments due to their low nuclear recoils. We have comprehensively explored production of hadrophilic light DM from cosmic ray collisions with the atmosphere, whose relativistic flux leads to observable experimental signatures. The decays of mesons resulting from collisions copiously source an irreducible flux of light DM for all terrestrial experiments. In this sense, our work applies to any hadrophilic scalars which couple to a new light neutral state, irrespective of whether or not that state accounts for the DM density. 

Additionally, for the first time we analyze DM production associated with $\eta$, $\eta^{\prime}$ as well as $K^+$ mesons. We implement an exact treatment of DM flux attenuation as it traverses the Earth towards an underground detector, finding that nuclear form factor effects can dramatically impact the viable DM parameter space. We set new limits on hadrophilic light DM from XENON1T experiment and establish sensitivity projections for future experiments, such as DARWIN which will cover new low mass DM and low mass mediator parameter space. Future extensions of this work may include extending the DM model to examine inelastic DM transitions, which will impact both the recoil spectra and the DM attenuation. Another future extension would be to examine DM flux contributions beyond meson decay, such as scalar bremsstrahlung~\cite{Batell:2018fqo} which may provide access to higher mass DM. 

\section*{Acknowledgments}

We dedicate this work to Tom Gaisser, whose contributions in cosmic-ray physics made this work (and many others!) possible. We thank Yu-Dai Tsai for discussions.
C.A.A. is supported by the Faculty of Arts and Sciences of Harvard University, and the Alfred P. Sloan Foundation.
I.M.S. is supported by the U.S. Department of Energy under the award number DE-SC0020250.
V.M. is supported by ANID-PCHA/DOCTORADO BECAS CHILE/2018-72180000.
V.T. is supported by the World Premier International Research Center Initiative (WPI), MEXT, Japan. 

\appendix
 
\section{Meson Decay Scalar Production}
\label{sec:scalarmed}

Sub-GeV scalar mediator $S$ can be in meson decays from atmospheric collisions through a variety of channels. The resulting leading branching ratios for $S$ can be found via chiral Lagrangian approach~\cite{Batell:2018fqo}:
\begin{align} \label{eq:branch}
{\rm Br}(\eta \rightarrow \pi^0 S) =&~\dfrac{C_{\eta}^2 g_u^2 B^2}{16 \pi m_{\eta} \Gamma_{\eta}} \lambda_{1/2}\Big(1, \dfrac{m_S^2}{m_{\eta}^2}, \dfrac{m_{\pi^0}^2}{m_{\eta}^2}\Big), \notag\\
    {\rm Br}(\eta^{\prime} \rightarrow \pi^0 S) =&~\dfrac{C_{\eta^{\prime}}^2 g_u^2 B^2}{16 \pi m_{\eta^{\prime}} \Gamma_{\eta^{\prime}}} \lambda_{1/2}\Big(1, \dfrac{m_S^2}{m_{\eta^{\prime}}^2}, \dfrac{m_{\pi^0}^2}{m_{\eta^{\prime}}^2}\Big), \notag\\
    {\rm Br}(K^{+} \rightarrow \pi^+ S) =&~\dfrac{G_F^2 g_u^2 f_{\pi}^2 f_K^2 B^2}{8 \pi m_{K^+} \Gamma_{K^{+}}} |V_{ud}V_{us}|^2  \\ 
    &~\times \lambda_{1/2}\Big(1, \dfrac{m_S^2}{m_{K^+}^2}, \dfrac{m_{\pi^+}^2}{m_{K^+}^2}\Big), \notag
\end{align}
where $B \simeq m_{\pi}^2/(m_u + m_d) \simeq 2.6$~GeV, $C_{\eta}(\eta^{\prime}) = \sqrt{1/3} \cos\theta^{\prime} \mp \sqrt{2/3} \sin \theta^{\prime}$ parametrize $\eta - \eta^{\prime}$ mixing with $\theta^{\prime} \simeq -20^{\circ}$, $\lambda(a,b,c) = a^2 + b^2 + c^2 - 2 a b - 2 b c - 2 a c$, $G_F = 1.167 \times 10^{-5}$~GeV$^{-2}$ is the Fermi coupling constant, $|V_{ud}|$ and $|V_{us}|$ are the CKM matrix elements, $f_{\pi}$ and $f_K$ are decay constants and $\Gamma_{\eta}$, $\Gamma_{\eta^{\prime}}$ and $\Gamma_{K^+}$ are the decay widths~\cite{Zyla:2020zbs}.
Throughout we consider that Br$(S \rightarrow \chi\chi) = 1$. 

\bibliography{biblio.bib}
\end{document}